\documentclass[english,aps,floats,onecolumn,showpacs,nofootinbib]{revtex4}
\usepackage{pslatex}
\usepackage[T1]{fontenc}
\usepackage[latin1]{inputenc}
\usepackage{graphicx}
\usepackage{epsfig}

\usepackage{calc}
\usepackage{ifthen}

{
{
{
\newcommand{\bea}{\begin{eqnarray}}
\newcommand{\eea}{\end{eqnarray}}

\newcommand{\nc}{\newcommand}
\nc{\renc}{\renewcommand}
\nc{\eqs}[2]{\mbox{Eqs.~(\ref{#1},\,\ref{#2})}}
\nc{\eq}[1]{\mbox{Eq.~(\ref{#1})}}
\nc{\figs}[2]{\mbox{Figs.~(\ref{#1},\,\ref{#2})}}
\nc{\fig}[1]{\mbox{Fig~.(\ref{#1})}}
\nc{\be}[1]{\begin{equation} \mbox{$\label{#1}$}}
\nc{\ee}{\vspace{0.1cm}\end{equation}}

\newcommand{\bean}{\begin{eqnarray*}}
\newcommand{\eean}{\end{eqnarray*}}

\def\GeV{{\rm \ GeV}}

\def\lae{\;^{<}_{\sim} \;} \def\gae{\; ^{>}_{\sim} \;}

\begin{document}
\title{A Minimal Sub-Planckian Axion Inflation Model with Large Tensor-to-Scalar Ratio}
\author{John McDonald}
\email{j.mcdonald@lancaster.ac.uk}
\affiliation{Dept. of Physics, University of 
Lancaster, Lancaster LA1 4YB, UK}

\begin{abstract}

   We present a minimal axion inflation model which can generate a large tensor-to-scalar ratio while remaining sub-Planckian. The modulus of a complex scalar field $\Phi$ with a $\lambda |\Phi|^4$ potential couples directly to the gauge field of a strongly-coupled sector via a term of the form $(|\Phi|/M_{Pl})^{m} F \tilde{F}$. This generates a minimum of the potential which is aperiodic in 
the phase. The resulting inflation model is equivalent to a $\phi^{4/(m+1)}$ chaotic inflation model. For the natural case of a leading-order portal-like interaction of the form $\Phi^{\dagger}\Phi F \tilde{F}$, the model is equivalent to a $\phi^{4/3}$ chaotic inflation model and predicts a tensor-to-scalar ratio $r = 16/3N = 0.097$ and a scalar spectral index $n_{s} = 1-5/3N = 0.970$. The value of $|\Phi|$ remains sub-Planckian throughout the observable era of inflation, with $|\Phi|  \lae 0.01 M_{Pl}$ for $N \lesssim 60$ when $\lambda \sim 1$.

\end{abstract}
 \maketitle

\section{Introduction}

      The recent observation of a large B-mode signal in the CMB by BICEP2 \cite{bicep1,bicep2} may be an indication of a large tensor-to-scalar ratio $r \sim 0.1$ from inflation. The existence of a large primordial B-mode signal will be confirmed or excluded in the near future by Planck. This has focused attention on how inflation can naturally generate a large value for $r$. The Lyth bound \cite{lythb, antusch} shows that this requires a super-Planckian value for the inflaton field in single-field inflation\footnote{$M_{Pl} = (8 \pi G)^{-1/2}$.}, with typically\footnote{In \cite{hertzberg}, it is shown that a hybrid inflation model based on unification energy with $\phi \sim \sqrt{8 \pi} M_{Pl}$ can fit observations.} $\phi \sim 10 M_{Pl}$. This is difficult to understand theoretically, as we expect new physics to arise at the Planck scale which would be expected to introduce non-renormalizable terms into the inflaton potential, excluding super-Planckian values for the inflaton. This is particularly true in supersymmetry, since $M_{Pl}$ is a fundamental scale of supergravity and the existence of higher-order terms in the K\"ahler potential and superpotential appears to be natural. The problem might be solved by imposing a shift-symmetry which eliminates the potential terms. This must then be slightly broken to permit an inflaton potential, as in the models discussed in \cite{shift} and \cite{shift2} \footnote{For an alternative shift-symmetry mechanism motivated by extra-dimensions, see \cite{kaloper}.}.

   Should such a "Planck barrier" exist, we will need to construct an inflation  model where the field values are sub-Planckian 
throughout. This can be done using a potential based on two or more fields, such that the inflaton trajectory can travel a super-Planckian distance while the fields remain sub-Planckian. This typically involves a spiralling trajectory in field space. 

  Models based on axion-like fields in string theory have been proposed to achieve this \cite{axmon,dantes}. (For a review of axion inflation models in general, see \cite{axinfrev}.) Alternatively, one can adopt a more direct approach to model building, introducing just what is necessary to produce sub-Planckian inflation. Models based on two axions with two anomalous gauge groups were proposed in \cite{peloso,nilles}.  In this paper we will present a minimal sub-Planckian axion inflation model, based on a single complex field $\Phi$ interacting with a single strongly-coupled gauge group.  We will show that in the case where all non-renormalizable interactions are suppressed by powers of the Planck scale, the most likely model is equivalent to a $\phi^{4/3}$ chaotic inflation model, with sub-Planckian values for $|\Phi|$ throughout the observable era of inflation.

\section{A minimal sub-Planckian axion inflation model with large $r$}

     The model is similar to the heavy quark axion (KSVZ) model \cite{hqa}. We introduce a complex field $\Phi$, which is a gauge singlet, and a chiral $U(1)_{A}$ global symmetry. We also introduce a heavy fermion $Q$ in the fundamental representation of the gauge group, where $\Phi$ interacts with $Q$ via
\be{e1} h \Phi \overline{Q}_{R} Q_{L} + h.c.    ~.\ee
The $\Phi$ scalar potential is 
\be{e2} V(\Phi) = -\mu^2 |\Phi|^2 + \lambda |\Phi|^4    ~,\ee
where $\lambda \sim 1$ is expected dimensionally. $Q$ gains its mass from the VEV of $\Phi$. During inflation we can assume that $|\Phi| \gg \mu$. Therefore we will set $\mu = 0$ in the following and consider $V(\Phi) = \lambda |\Phi|^4$. $\Phi$ has charge $+1$ under $U(1)_{A}$ and $Q$ has charge $+1/2$. The phase $\theta$ of $\Phi \; (= \phi e^{i \theta}/\sqrt{2})$ in \eq{e1} can be rotated away via a local chiral transformation of the fermions. This results in a $U(1)_{A}$-breaking interaction of $\theta$ with the gauge fields due to the chiral anomaly,  
\be{e3} \frac{g^2 \theta}{32 \pi^2} F \tilde{F}    ~.\ee
Here $F\tilde{F} = F_{\mu \nu} \tilde{F}^{\mu \nu}$, $\tilde{F}^{\mu \nu} = \frac{1}{2} \epsilon^{\mu\nu\rho\sigma} F_{\rho \sigma}$ and $g$ is the gauge coupling of the strongly-coupled gauge sector. (Gauge indices are suppressed.) 
In general, we can also include a $U(1)_{A}$-symmetric non-renormalizable interaction of the form 
\be{e4}  g^2 \xi \frac{|\Phi|^{m}}{M_{Pl}^m} F \tilde{F}  ~,\ee
where $\xi$ is a dimensionless parameter.  
The combination of \eq{e3} and \eq{e4} can then be written as 
\be{e5}  \frac{g^2}{32 \pi^2} \left( \frac{|\Phi|^m}{\Lambda^m} + \theta \right) F \tilde{F}   ~,\ee
where $\Lambda = M_{Pl}/(32 \pi^{2} \xi)^{1/m}$.  We define the strong coupling scale to be 
$\Lambda_{sc}$. The potential term generated by the strongly-coupled gauge sector is then 
\be{e6}  V_{sc}(|\Phi|,\theta) = -\Lambda_{sc}^{4} \cos \left( \frac{|\Phi|^m}{\Lambda^m} + \theta \right)   ~.\ee
Therefore we can define the full potential for $\Phi$ during inflation to be  
\be{e7}  V_{tot}(\Phi) = V(\Phi) + V_{sc}(\Phi) + \Lambda_{sc}^{4} = \lambda|\Phi|^4 + \Lambda_{sc}^{4} \left[1 - \cos \left( \frac{|\Phi|^m}{\Lambda^m} + \theta \right) \right]   ~,\ee
where we have added a constant term $\Lambda_{sc}^4$ so that the potential equals zero at the global minimum\footnote{
In \cite{dantes}, a string-motivated two-axion model is considered which is equivalent \eq{e7} with $m=1$ and $V(|\Phi|)$ proportional to a general power of $|\Phi|$. Axion monodromy favours $V(\Phi) \propto |\Phi|$ \cite{axmon}.}. 

This potential has a minimum which is aperiodic in $\theta$. Along the $\phi$ direction for a given $\theta$, the strong-coupling term modulates the $|\Phi|^4$ potential. For a range of parameters which we will determine below, there are local minimum of the potential as a function of $|\Phi|$, which correspond to the cosine term being close to 1. The value of $|\Phi|$ at these minima satisfies
\be{e8} \frac{|\Phi|^{m}}{\Lambda^{m}}  \approx 2 n \pi - \theta ~,\ee
where $n$ is an integer. This results in a spiralling groove inscribed on the $|\Phi|^4$ potential in the complex $\Phi$ plane. Inflation can occur along this groove, allowing the field to traverse a long distance in field space while $|\Phi|$ remains sub-Planckian throughout.

  The inflation dynamics of this model are the same as that of the model presented in \cite{mod1}, where a multiplicative modulation of the $|\Phi|^4$ potential was considered.  The distance $a$ along the minimum in field space is related to $\theta$ by 
\be{e9}  da = \sqrt{\phi^{2} + \left(\frac{d\phi}{d\theta}\right)^{2}  } d \theta  ~,\ee From \eq{e8}, 
\be{e10} \frac{d \phi}{d \theta} = -\left( \frac{\sqrt{2} \Lambda}{\phi} \right)^{m} \frac{\phi}{m}   ~.\ee
If $(\sqrt{2}\Lambda/\phi)^{2m} \ll 1$ then to a good approximation $da = \phi(\theta) d\theta$. In this case we can consider $a$ to be a canonically normalized field along the minimum of the potential. The model will behave as a single field inflation model if the field $\phi$ orthogonal to $a$ has a mass much larger than $H$. (We  will derive the condition for this to be true later.) Using $da = \phi(\theta) d\theta$, we find 
\be{e11}  \int da = \int \phi \left( \frac{d \theta}{d \phi} \right) d \phi  = \int -m \left(\frac{\phi}{\sqrt{2} \Lambda} \right)^{m} d \phi  
 ~.\ee
Therefore
\be{e12}  \phi = 
\left( \left(\sqrt{2} \Lambda\right)^{m} \left(\frac{m+1}{m}\right) \right)^{\frac{1}{m+1}} \left( \left(\frac{m}{m+1}\right) \frac{\phi_{0}^{m+1}}{\left(\sqrt{2} \Lambda\right)^{m} } - a\right)^{\frac{1}{m+1}}    ~,\ee
where we have defined $a = 0$ at $\phi = \phi_{0}$. 
We can then define a new slow-roll field, $\hat{a}$, given by
\be{e13} \hat{a} = \frac{m}{\left(m+1\right)} \frac{\phi^{m+1}}{\left(\sqrt{2} \Lambda\right)^{m} }  ~.\ee 
Along the minimum ("groove"), the potential is  
\be{e14} V(\hat{a}) = \frac{\lambda \phi^{4}(\hat{a})}{4} = \frac{\lambda}{4} 
\left( \left(\sqrt{2} \Lambda\right)^{m} \left(\frac{m+1}{m}\right) \right)^{\frac{4}{m+1}} \hat{a}^{\frac{4}{m+1}}    ~.\ee 
The model will therefore have the same inflaton dynamics as a $\phi^{4/(m+1)}$ chaotic inflation model.  The spectral index $n_{s}$ and the tensor-to-scalar ratio $r$ as a function of the number of e-foldings $N$ are therefore
\be{e15} n_{s} =  1 - \left(\frac{m+3}{m+1} \right) \frac{1}{N}    ~\ee 
and
\be{e16} r = \left(\frac{16}{m+1}\right) \frac{1}{N}  ~.\ee
$\hat{a}$ and $\phi$ are related to $N$ by 
\be{e17}  \frac{\hat{a}}{M_{Pl}} = \left(\frac{8N}{m+1}\right)^{1/2}  ~\ee
and 
\be{e18} \left(\frac{\phi}{\sqrt{2} \Lambda}\right)^{m+1}
= \left(\frac{4\left(m+1\right)N}{m^2} \right)^{1/2}  \frac{M_{Pl}}{\Lambda}  ~.\ee

    A case of particular interest is where the field $\Phi$ is the fundamental object out of which the effective theory is constructed, by which we mean that only integer powers of $\Phi$ and $\Phi^{\dagger}$ occur in the effective theory at low energies. This excludes, for example, terms proportional to $|\Phi|$. In this case the natural $U(1)_{A}$-invariant combination is the bilinear $\Phi^{\dagger}\Phi$. If Planck-scale physics generates a $U(1)_{A}$-symmetric interaction of the form $g^2 \xi f(\Phi^{\dagger}\Phi/M_{Pl}^{2})F \tilde{F}$, where $f(x)$ is assumed to be expandable with a leading-order term proportional to $x$ when $x \ll 1$,  there will be a portal-like leading-order interaction of the form $\Phi^{\dagger}\Phi F \tilde{F}$, corresponding to \eq{e4} with $m = 2$.
      In the following we will focus attention on the $m = 2$ model, which we consider to be the most likely form of coupling to the gauge sector. 

     The $m = 2$ model is equivalent to a $\phi^{4/3}$  
chaotic inflation model. In this case the predictions for $N =55$ are $n_{s} = 1 -5/3N = 0.970$ and $r = 16/3N = 0.097$, with a negligible running of $n_{s}$. The spectral index is in reasonable agreement with the value determined by Planck, $n_{s} = 0.9624 \pm 0.0075$ (Planck + WP, assuming negligible running and including a tensor component \cite{planckinf}).   The values $|\Phi|$ and $\Lambda$ are determined by the curvature perturbation power spectrum, 
\be{e19} {\cal P}_{\zeta} = \frac{V}{24 \pi^2 \epsilon M_{Pl}^{4}}    ~,\ee
where $P_{\zeta}^{1/2} = 4.8 \times 10^{-5}$.
Using \eq{e14} and \eq{e18} we obtain
\be{e19a}  \frac{\Lambda}{M_{Pl}}  =  \left( \frac{24 \pi^2 P_{\zeta}}{(m+1) \lambda } \right)^{\frac{m+1}{4m}} 
\left( \frac{m^2}{4(m+1)} \right)^{\frac{1}{2m}} \left( \frac{1}{N} \right)^{\frac{m+3}{4m}}    ~\ee
and 
\be{e21} \frac{|\Phi|}{M_{Pl}} = \left( \frac{4 (m+1) N}{m^{2}} \right)^{\frac{1}{2(m+1)}} \left(  \frac{\Lambda}{M_{Pl}} \right)^{\frac{m}{m+1}}   ~.\ee
For $m = 2$ and $N = 55$ these become
\be{e20} \frac{\Lambda}{M_{Pl}} = \left( \frac{8 \pi^2 P_{\zeta}}{\lambda} \right)^{3/8} \left( \frac{1}{3} \right)^{1/4} 
\left( \frac{1}{N} \right)^{5/8} = 1.8 \times 10^{-4}\; \lambda^{-3/8}   ~\ee
and 
\be{e22}  \frac{|\Phi|}{M_{Pl}} = \left(3 N\right)^{1/6} \left( \frac{\Lambda}{M_{Pl}} \right)^{2/3}  = 0.0075 \; \lambda^{-1/4}   ~.\ee
Thus we find that $|\Phi| \lae 0.01 M_{Pl}$ throughout the observable era of inflation when $\lambda \sim 1$. Therefore the model can produce a large value for the tensor-to-scalar ratio while remaining sub-Planckian throughout. The model also allows conventional particle physics strength $|\Phi|^4$ potentials with $\lambda \sim 1$ to serve as basis for the inflaton potential.

The value of $\Lambda$ is small compared with $M_{Pl}$ when $\lambda \sim 1$. This means that the dimensionless coupling $\xi$ is necessarily large\footnote{Such large dimensionless couplings are not without precedent in inflation models. In non-minimally coupled models of inflation based on a $|\Phi|^4$ potential \cite{salopek},  
the coupling $\xi$ between $\Phi^{\dagger}\Phi$ and the Ricci scalar $R$ is given by $\xi \approx  10^{5} \sqrt{\lambda}$ \cite{salopek}. In both models, the large value of $\xi$ effectively replaces the small scalar coupling of conventional $\phi^4$ chaotic inflation models.} 
 when $\lambda \sim 1$, 
\be{e23} \xi = \frac{M_{Pl}^2}{32 \pi^2 \Lambda^2} = 9.8 \times 10^{4} \; \lambda^{3/4}     ~.\ee

  The model is still significantly sub-Planckian even when $\xi \sim 1$. To show this, we can express $|\Phi|/M_{Pl}$ and $\lambda$ as functions of $\xi$,
\be{e22a} \frac{|\Phi|}{M_{Pl}} = 0.35 \; \xi^{-1/3}    ~\ee
and 
\be{e22b} \lambda = 2.2 \times 10^{-7} \; \xi^{4/3}    ~.\ee 
Therefore $\xi = 1$ implies that $|\Phi|/M_{Pl} = 0.35$. 
However, if we expect the leading-order Planck correction to the potential to be of the order of $|\Phi|^6/M_{Pl}^2$, then this will dominate the $\lambda |\Phi|^4$ term due to the small value of $\lambda$. On the other hand, if we simply wish $|\Phi|$ to be sub-Planckian so that the potential is calculable with respect to Planck corrections (for example if $V(|\Phi|) = \lambda v(|\Phi|^{2}/M_{pl}^{2})$, where $v(x)$ can be expanded in $x$), then $\xi \sim 1$ is possible. We note that in this case the model with $\xi \sim 1$ can achieve a sub-Planckian  $|\Phi|$ with far fewer new fields than in the case of N-flation, where, in order to have $\phi_{n} < \sqrt{2} \times 0.35 M_{Pl}$, it is necessary to have $N > 900$ scalar fields \cite{nflation}.

 We next determine the conditions for the underlying assumptions of the model to be consistent. We have assumed that a local minimum in the radial direction exists. This requires that the derivative of the potential in the $\phi$ direction is dominated by the strong-coupling term, $V^{'}_{sc}(\phi) \gg V^{'}(\phi)$. This imposes a lower bound on $\Lambda_{sc}$,  
\be{e24} \Lambda_{sc}^{4} \gg \frac{\sqrt{2}^{\; m} \lambda  \Lambda^{m} \phi^{4 -m}}{m}   ~.\ee
For $m = 2$ this becomes 
\be{e25} \Lambda_{sc} > \lambda^{1/4} (\Lambda \phi)^{1/2} ~.\ee
Using \eq{e20} and \eq{e22} we obtain
\be{e26}   \Lambda_{sc} > 1.4 \times 10^{-3} \lambda^{-1/16} M_{Pl}  ~.\ee 
We have also assumed that the effective mass squared at the minimum in the radial direction, which is dominated by the strong coupling term,  is large enough to reduce the dynamics of the model to a single-field inflation model in the $\hat{a}$ direction. This requires that $V^{''}_{sc}(\phi) \gg H^2$ at the minimum. This also imposes a lower bound on $\Lambda_{sc}$, 
\be{e27} \Lambda_{sc}^{4} \gg \frac{\lambda}{12} \frac{\left(\sqrt{2} \Lambda\right)^{2m}}{m^2} \frac{\phi^{6-2m}}{M_{Pl}^{2}}    ~.\ee
For $m = 2$ this becomes, 
\be{e28}  \Lambda_{sc} > \left( \frac{\lambda}{12} \right)^{1/4} \left( \frac{\phi}{M_{Pl}}\right)^{1/2} \Lambda ~.\ee
Using \eq{e20} and \eq{e22} we obtain
\be{e30} \Lambda_{sc} > 1.0 \times 10^{-5} \; \lambda^{-1/4}  M_{Pl}  ~.\ee
This is a weaker lower bound than that from the existence of the minimum, \eq{e26}. 
Therefore if $\Lambda_{sc} > 3.4 \times 10^{15} \lambda^{-1/16} \GeV$ then the $m = 2$ model can consistently account for sub-Planckian inflation while generating a large value for $r$. The assumptions underlying the model with $\lambda \sim 1$ will therefore be well-satisfied if\footnote{We note that these values of $\Lambda_{sc}$ are much larger than the value of $H$ at $N = 55$ ($ = 8 \times 10^{13} \lambda^{1/2} \GeV)$. Since $H$ characterises the energy of the inflationary fluctuations, the assumption of strong coupling is consistent.} $\Lambda_{sc} \gae 10^{16} \GeV$.

\section{Conclusions}

     We have presented a minimal axion inflation model which is consistent with the observed value of $n_{s}$ and which can generate a large tensor-to-scalar ratio $r \sim 0.1$ while remaining sub-Planckian throughout. The model also allows $\lambda |\Phi|^4$ potentials with $\lambda \sim 1$ to serve as basis for the inflaton potential.
The model requires only a single complex field and a single strongly-coupled gauge group.
We emphasize that the model is constructed as a straightforward particle physics model, similar in construction to the KSVZ axion model, with the only new element being the direct coupling of the scalar field to the $F \tilde{F}$ term. This contrasts with models where a specific ultra-violet completion is required, in particular string theory, and where essential dynamics is outwith the sub-Planckian four dimensional effective field theory. For the case where the effective theory at low energies is constructed from integer powers of $\Phi$ and $\Phi^{\dagger}$ and all non-renormalizable terms are part of an expansion in inverse powers of the Planck scale, the most likely coupling of $\Phi$ to the gauge sector has the portal-like form $\Phi^{\dagger} \Phi F \tilde{F}$. In this case the model is dynamically equivalent to a $\phi^{4/3}$ chaotic inflation model, with $n_{s} = 0.970$ and $r = 0.097$. 
While models of chaotic inflation with fractional powers have been proposed previously, it is unusual and significant that the specific power $n = 4/3$ is favoured by the model presented here. The model is explicitly sub-Planckian throughout the observable era of inflation, with $|\Phi| \lae 0.01 M_{Pl}$ for $N \lae 60$. The strong coupling scale must be greater than $10^{16} \GeV$ for the model to be consistent. If the $\lambda |\Phi|^{4}$ coupling takes the dimensionally natural value expected in conventional particle physics theories, $\lambda \sim 1$, then the dimensionless coupling $\xi$ of $\Phi^{\dagger}\Phi$ to the gauge fields must be large, $\xi \sim 10^{5}$, in order to reproduce the observed CMB temperature anisotropies. Values of $\xi \sim 1$ also result in sub-Planckian $|\Phi|$, with $|\Phi|/M_{Pl} = 0.35 \; \xi^{-1/3} M_{Pl}$. This case requires additional suppression of Planck corrections to the potential but allows the potential to be calculable with respect to such corrections.

\section*{Acknowledgements} This work was partially supported by STFC grant
ST/J000418/1.

\renewcommand{\theequation}{A-\arabic{equation}}
 % redefine the command that creates the equation no.
 \setcounter{equation}{0} 
% reset counter 

\end{document}